\newif\ifJournal
\Journalfalse         %

\ifJournal

 \documentclass[secthm]{elsart}

\else

  \documentclass[11pt]{article}

\fi

\usepackage{chicagor}

\usepackage{amsmath}
\allowdisplaybreaks[2]

\usepackage{amssymb}
\usepackage{times}
\usepackage{url}

\newif\ifpdf
\ifx\pdfoutput\undefined
  \pdffalse
\else
  \pdfoutput=1
  \pdftrue
\fi

 \ifpdf
  \usepackage[pdftex]{graphicx}
  \DeclareGraphicsExtensions{.pdf}
 \else
   \usepackage{graphicx}
   \DeclareGraphicsExtensions{.eps}
 \fi

\ifJournal

  \newenvironment{theorem}{\begin{thm}}{\end{thm}}
  \newenvironment{corollary}{\begin{cor}}{\end{cor}}
  \newenvironment{proposition}{\begin{prop}}{\end{prop}}
  \newenvironment{prf}{\begin{pf}}{\end{pf}}

  \newenvironment{example}{\begin{exmp}}{\end{exmp}}

\else

\newcommand{\thmcolon}{\hspace{-.55em}{\bf :}}
\newtheorem{THEOREM}{Theorem}[section]
\newenvironment{theorem}{\begin{THEOREM} \thmcolon  }%
                        {\end{THEOREM}}
\newtheorem{LEMMA}[THEOREM]{Lemma}
                      {\end{LEMMA}}
\newtheorem{COROLLARY}[THEOREM]{Corollary}
\newenvironment{corollary}{\begin{COROLLARY} \thmcolon  }%
                          {\end{COROLLARY}}
\newtheorem{PROPOSITION}[THEOREM]{Proposition}
\newenvironment{proposition}{\begin{PROPOSITION} \thmcolon  }%
                            {\end{PROPOSITION}}
\newtheorem{DEFINITION}[THEOREM]{Definition}
                            { \bbox\end{DEFINITION}}
\newenvironment{definition*}{\begin{DEFINITION} \thmcolon  \rm}%
                            {\end{DEFINITION}}
\newtheorem{CLAIM}[THEOREM]{Claim}
                            {\end{CLAIM}}
\newtheorem{EXAMPLE}[THEOREM]{Example}
\newenvironment{example}{\begin{EXAMPLE} \thmcolon  \rm}%
                            { \bbox\end{EXAMPLE}}
\newenvironment{example*}{\begin{EXAMPLE} \thmcolon  \rm}%
                            {\end{EXAMPLE}}
\newtheorem{REMARK}[THEOREM]{Remark}
                            { \bbox\end{REMARK}}
\newenvironment{remark*}{\begin{REMARK} \thmcolon  \rm}%
                            {\end{REMARK}}

\newenvironment{prf}{\noindent{\bf Proof:} }{ \vspace{0.1in}}

\newcommand{\bbox}{\vrule height7pt width4pt depth1pt}
 \newcommand{\qed}{\bbox}

\fi

\newcommand{\primprogs}{\ensuremath{\mathcal{P}}}
\newcommand{\primtests}{\ensuremath{\mathcal{B}}}

\newcommand{\langs}{\ensuremath{\mathcal{L}}}

\newcommand{\Neg}[1]{\ensuremath{\overline{#1}}}

\newcommand{\KA}{\ensuremath{\mathsf{KA}}}
\newcommand{\KAT}{\ensuremath{\mathsf{KAT}}}

\newcommand{\<}{\langle}
\renewcommand{\>}{\rangle}
\renewcommand{\emptyset}{\varnothing}
\newcommand{\trans}[1]{\stackrel{#1}{\longrightarrow}}
\newcommand{\lit}{\mathit{lit}}
\newcommand{\base}{\mathit{base}}
\newcommand{\MA}{\mathcal{MA}}

\newcommand{\hD}{\ensuremath{\hat{D}}}
\newcommand{\hT}{\ensuremath{\hat{T}}}
\newcommand{\hepsilon}{\ensuremath{\hat{\epsilon}}}
\newcommand{\eqACI}{\stackrel{\scriptscriptstyle\rm ACI}{=}}
\newcommand{\hR}{\ensuremath{\hat{R}}}
\newcommand{\hRACI}{\hR^{\scriptscriptstyle\rm ACI}}

\newcommand{\Rule}[2]{          %
  \begin{array}{c}
  #1 \\\hline
  #2
  \end{array}}

\def\condarray#1#2#3{
    \left\{\begin{array}{ll}
        #2 & \mbox{if $#1$} \\
        #3 & \mbox{otherwise}
    \end{array}\right.}

\newcommand{\condition}[1]{\textbf{#1}}

\newenvironment{prog}{\begin{array}[t]{@{}l@{}}}{\end{array}}

\newcommand{\COMMENTOUT}[1]{}

\begin{document}

\ifJournal

\begin{frontmatter}
  \title{A Coalgebraic Approach to Kleene Algebra with Tests\thanksref{prelim}}
  \author{Hubie Chen}
  \ead{hubes@cs.cornell.edu} and
  \author{Riccardo Pucella}
  \ead{riccardo@cs.cornell.edu}
  \address{Department of Computer Science, Cornell University, Ithaca, NY 14853 USA}

\thanks[prelim]{A preliminary version of this paper appeared in the
\emph{Proceedings of the Sixth International Workshop on Coalgebraic Methods
in Computer Science}, Electronic Notes in Theoretical Computer
Science, Volume 82.1, 2003.}

\else

  \title{A Coalgebraic Approach to Kleene Algebra with
Tests\thanks{This paper is essentially the same as one that will
appear in \emph{Theoretical Computer Science}. A preliminary version appeared in the
\emph{Proceedings of the Sixth International Workshop on Coalgebraic Methods
in Computer Science}, Electronic Notes in Theoretical Computer
Science, Volume 82.1, 2003.}}
  \author{Hubie Chen\\
  Cornell University\\
  Ithaca, NY 14853\\
  hubes@cs.cornell.edu
  \and 
  Riccardo Pucella\\
  Cornell University\\
  Ithaca, NY 14853\\
  riccardo@cs.cornell.edu} 
  \date{} 

  \maketitle

\fi

\begin{abstract}
Kleene algebra with tests is an extension of Kleene algebra, the
algebra of regular expressions, which can be used to reason about
programs. We develop a coalgebraic theory of Kleene algebra with
Tests, along the lines of the coalgebraic theory of regular
expressions based on deterministic automata. Since the known
automata-theoretic presentation of Kleene algebra with tests does not
lend itself to a coalgebraic theory, we define a new interpretation of
Kleene algebra with tests expressions and a corresponding
automata-theoretic presentation. One outcome of the theory is a
coinductive proof principle, that can be used to establish equivalence
of our Kleene algebra with tests expressions.
\end{abstract}

\ifJournal \end{frontmatter} \fi

\section{Introduction}

Kleene algebra (\KA) is the algebra of regular expressions
\cite{Conway71,Kleene56}.  As is well known, the theory of regular
expressions enjoys a strong connection with the theory of finite-state
automata. This connection was used by Rutten \citeyear{Rutten98} to
give a coalgebraic treatment of regular expressions.  One of the
fruits of this coalgebraic treatment is \emph{coinduction}, a proof
technique for demonstrating the equivalence of regular expressions
\cite{Rutten00}.  Other methods for proving the equality of regular
expressions have previously been established---for instance,
reasoning by using a sound and complete axiomatization
\cite{Kozen94,Salomaa66}, or by minimization of automata representing the
expressions \cite{Hopcroft79}. However, the coinduction proof technique
can give relatively short proofs, and is fairly simple to apply.

Recently, Kozen \citeyear{Kozen97} introduced Kleene algebra with
tests (\KAT), an extension of $\KA$ designed for the particular
purpose of reasoning about programs and their properties.  The regular
expressions of $\KAT$ allow one to intersperse boolean tests along
with program actions, permitting the convenient modelling of
programming constructs such as conditionals and \emph{while} loops.
The utility of $\KAT$ is evidenced by the fact that it subsumes
propositional Hoare logic, providing a complete deductive system for
Hoare-style inference rules for partial correctness assertions
\cite{Kozen99}.

The goal of this paper is to develop a coalgebraic theory of $\KAT$,
paralleling the coalgebraic treatment of $\KA$.  Our coalgebraic
theory yields a coinductive proof principle for demonstrating the
equality of $\KAT$ expressions, in analogy to the coinductive proof
principle for regular expressions.  The development of our coalgebraic
theory proceeds as follows.  We first introduce a form of
deterministic automaton and define the language accepted by such an
automaton.  Next, we develop the theory of such automata, showing that
coinduction can be applied to the class of languages representable by
our automata.  We then give a class of expressions, which play the
same role as the regular expressions in classical automata theory, and
fairly simple rules for computing derivatives of these expressions.

The difficulty of our endeavor is that the known automata-theoretic
presentation of $\KAT$ \cite{r:kozen03} does not lend itself to a
coalgebraic theory. Moreover, the notion of derivative, essential to
the coinduction proof principle in this context, is not readily
definable for $\KAT$ expressions as they are defined by Kozen
\citeyear{Kozen97}.  Roughly, these difficulties arise from tests
being commutative and idempotent, and suggest that tests need to be
handled in a special way.  In order for the coalgebraic theory to
interact smoothly with tests, we introduce a \emph{type system} along
with new notions of strings, languages, automata, and expressions,
which we call \emph{mixed strings}, \emph{mixed languages},
\emph{mixed automata}, and \emph{mixed expressions}, respectively. (We
note that none of these new notions coincide with those already
developed in the theory of $\KAT$.) All well-formed instances of these
notions can be assigned types by our type system.  Our type system is
inspired by the type system devised by Kozen
\citeyear{Kozen98,Kozen02} for $\KA$ and $\KAT$, but is designed to
address different issues.

This paper is structured as follows. In the next section, we introduce
 mixed strings and mixed languages, which will be used to
interpret our mixed expressions. In Section~\ref{s:automata},
we define a notion of mixed automaton that is used to accept mixed
languages. We then impose a coalgebraic structure on such automata. 
In Section~\ref{s:pseudo}, we introduce a sufficient
condition for proving equivalence that is more convenient than the
condition that we derive in Section~\ref{s:automata}. 
In
Section~\ref{s:expressions}, we introduce our type system for $\KAT$,
and connect typed $\KAT$ expressions with the mixed language they
accept.
In Section~\ref{s:example}, we give an example of how to use the
coalgebraic theory, via the coinductive proof principle, to establish
equivalence of typed $\KAT$ expressions.
In Section~\ref{s:completeness}, we show that our technique is
complete, that is, it can establish the equivalence of any two typed
$\KAT$ expressions that are in fact equivalent. 
We conclude in Section~\ref{s:conclusions} with considerations of
future work. 
\section{Mixed Languages}\label{s:languages}

In this section, we define the notions of mixed strings and mixed
languages that we will use throughout the paper. 
Mixed strings are a variant of the guarded strings introduced by
Kaplan \citeyear{r:kaplan69} as an abstract interpretation for program 
schemes; sets of guarded strings were used by Kozen
\citeyear{r:kozen03} as canonical models for Kleene algebra with
tests. Roughly speaking, a guarded string can be understood as a
computation where atomic actions are executed amidst the checking of
conditions, in the form of boolean tests. 
Mixed strings will be used as an interpretation for the mixed
expressions we introduce in Section~\ref{s:expressions}. 

Mixed strings are
defined over two alphabets: a set of primitive programs (denoted
$\primprogs$) and a set of primitive tests (denoted $\primtests$). We
allow $\primprogs$ to be infinite, but require that $\primtests$ be
finite. 
(We will see in Section~\ref{s:automata} where this finiteness
assumption comes in. Intuitively, this is because our automata will
process each primitive test individually.)
Primitive tests can be put together to form more complicated tests. A
\emph{literal} $l$ is a primitive test $b \in \primtests$ or its
negation $\Neg{b}$; the underlying primitive test $b$ is said to be
the base of the literal, and is denoted by $\base(l)$.  When $A$ is a
subset of $\primtests$, 
$\lit(A)$ denotes the set of all literals over
$A$. A \emph{test} is a nonempty set of literals with distinct
bases. Intuitively, a test can be understood as the conjunction of the
literals it comprises. 
The \emph{base of a test} $t$, denoted by $\base(t)$, is defined to be
the set $\{ \base(l) : l \in t \}$, in other words, the primitive
tests the test $t$ is made up from. 
We extend the notion of base to primitive programs, by defining the
\emph{base of a primitive program} $p \in \primprogs$ as $\emptyset$.

\begin{example}
Let $\primprogs=\{p,q\}$, and $\primtests=\{b,c,d\}$. The literals
$\lit(\primtests)$ of $\primtests$ are
$\{b,\Neg{b},c,\Neg{c},d,\Neg{d}\}$. Tests include $\{b,\Neg{c},d\}$ and
$\{\Neg{b},\Neg{d}\}$, but $\{b,\Neg{b},c\}$ is \emph{not} a test, as
$b$ and $\Neg{b}$ have the same base $b$. The base of
$\{b,\Neg{c},d\}$ is $\{b,c,d\}$. 
\end{example}

Primitive programs and tests are used to create mixed strings. 
A \emph{mixed string} is 
either the empty string, denoted by $\epsilon$,
or
a sequence
$\sigma = a_1 \ldots a_n$ (where $n \geq 1$) with the following
properties:
\begin{enumerate}
\item each $a_i$ is either a test or primitive program,
\item for $i = 1, \ldots, n-1$,
if $a_i$ is a test, then $a_{i+1}$ is a primitive program, 
\item for $i = 1, \ldots, n-1$,
if $a_i$ is a primitive program, then $a_{i+1}$ is a test,
 and 
\item for $i = 2, \ldots, n-1$, if $a_i$
is a test, then $\base(a_i) = \primtests$.
\end{enumerate}
Hence, a mixed string is an alternating sequence of primitive programs
and tests, where each test in the sequence is a ``complete'' test,
except possibly if it occurs as the first or the last element of the
sequence. 
This allows us to manipulate mixed strings on a finer level of
granularity; we can remove literals from the beginning
of a mixed strings and still obtain a mixed string. 
The length of the empty mixed string $\epsilon$ is $0$,
while the length of a mixed string $a_1\ldots a_n$ is $n$.

\begin{example}
Let $\primprogs=\{p,q\}$, and $\primtests=\{b,c,d\}$. Mixed strings
include $\epsilon$ (of length 0), $\{b\}$ and $p$ (both of length 1),
and $\{b\}p\{b,\Neg{c},d\}q\{\Neg{d}\}$ (of length 5). The sequence
$\{b\}p\{b,d\}q\{\Neg{d}\}$ is \emph{not} a mixed string, since
$\base(\{b,d\})\ne\primtests$.
\end{example}

We define the concatenation of two mixed strings $\sigma$ and
$\sigma'$, denoted by $\sigma \cdot \sigma'$, as follows.
If one of $\sigma, \sigma'$ is the empty string, then 
their concatenation is the other string.
If both
$\sigma  = a_1 \ldots a_n$ and
$\sigma' = b_1 \ldots b_m$
have non-zero length,
their concatenation is defined as:
\begin{enumerate}
\item $\tau = a_1 \ldots a_n b_1 \ldots b_m$ if 
exactly one of $a_n, b_1$ is a primitive program and 
$\tau$ is a mixed string;
\item $\tau = a_1 \ldots a_{n-1} (a_n \cup b_1) b_2 \ldots b_m$ if
$a_n$ and $b_1$ are tests such that 
$\base(a_n) \cap \base(b_1) = \emptyset$ and
$\tau$ is a mixed string; and is
\item  undefined otherwise.
\end{enumerate}
Intuitively, concatenation of the two strings is obtained by
concatenating the sequence of string elements, possibly by combining
the last test of the first string with the first test of the second
string, provided that the result is a valid mixed string. We note
that concatenation of strings is an associative operation.

\begin{example}
Let $\primprogs=\{p,q\}$, and $\primtests=\{b,c,d\}$. The
concatenation of the mixed strings $p$ and $\{b,c,d\}q$ is
$p\{b,c,d\}q$. Similarly, the concatenation of the mixed strings
$\{b\}p\{b,\Neg{c}\}$ and $\{d\}q\{\Neg{d}\}$ is the mixed string
$\{b\}p\{b,\Neg{c},d\}q\{\Neg{d}\}$. However, the concatenation of
$\{b\}p\{b,\Neg{c}\}$ and $\{b,d\}q$ is not defined, as
$\{b,\Neg{c}\}\cap\{b,d\}\ne\emptyset$. The concatenation
of $\{b\}p\{b,\Neg{c}\}$ and $q$ is also not defined, as
$\base(\{b,\Neg{c}\})\ne\primtests$, and thus $\{b\}p\{b,\Neg{c}\}q$
is not a mixed string.
\end{example}

We assign one or more types to mixed strings in the following way. A
type is of the form $A\rightarrow B$, where $A$ and $B$ are subsets of
$\primtests$. 
Intuitively, a mixed string has type $A\rightarrow B$ if the first
element of the string has base $A$, and it can be concatenated with an
element with base $B$. 
It will be the case that a mixed string of type $A\rightarrow B$ can
be concatenated with a mixed string of type $B\rightarrow C$ to obtain
a mixed string of type $A\rightarrow C$.

The mixed string $\epsilon$ has many
types, namely it has type $A \rightarrow A$, for all $A \in
\wp(\primtests)$.  A mixed string of length $1$ consisting of a single
test $t$ has type $\base(t) \cup A \rightarrow A$, for any $A \in
\wp(\primtests)$ such that $A \cap \base(t) = \emptyset$.  A mixed
string of length $1$ consisting of a single program $p$ has type
$\emptyset\rightarrow\primtests$. A mixed
string $a_1 \ldots a_n$ of length $n > 1$ has type
$\base(a_1) \rightarrow \primtests \setminus \base(a_n)$. 

\begin{example}
Let $\primprogs=\{p,q\}$, and $\primtests=\{b,c,d\}$. The mixed string 
$p\{b,\Neg{c},d\}$ has type $\emptyset\rightarrow\emptyset$. 
The mixed string $\{\Neg{d}\}p$ has type $\{d\}\rightarrow\primtests$. 
The mixed string $\{ b \} p \{ b,\Neg{c}, d \} q \{ b, \Neg{c}
\}$ has type $\{b\}\rightarrow\{d\}$. The concatenation of
$\{b\}p\{b,\Neg{c},d\}q\{b,\Neg{c}\}$ and $\{\Neg{d}\}p$, namely
$\{b\}p\{b,\Neg{c},d\}q\{b,\Neg{c},\Neg{d}\}p$, has type
$\{b\}\rightarrow\primtests$. 
\end{example}

A \emph{mixed language} is a set of mixed strings, and is
typeable, with type $A \rightarrow B$, if all of the mixed strings
it contains have type $A \rightarrow B$.
In this paper, we will only be concerned with typeable mixed
languages. 

We will be interested in different operations on mixed languages in
the following sections. When $L_1$,$L_2$, and $L$ are mixed languages,
we use the notation $L_1 \cdot L_2$ to denote the set $\{ \sigma_1
\cdot \sigma_2: \sigma_1 \in L_1, \sigma_2 \in L_2 \}$, $L^0$ to
denote the set $\{ \epsilon \}$, and for $n \geq 1$, $L^n$ to denote
the set $L \cdot L^{n-1}$. The following two operations will be useful
in Section~\ref{s:expressions}. The operator $T$, defined by
\[T(L)=\{ \sigma: \sigma \in L, |\sigma| = 1, \sigma \mbox{ is a test}\}\] 
extracts from a language all the mixed strings made up of a single
test. The operator $\epsilon$, defined by
\[\epsilon(L) = L \cap \{ \epsilon \}\]
essentially checks if the empty mixed string $\epsilon$ is in $L$,
since $\epsilon(L)$ is nonempty if and only if the empty mixed string
is in $L$.

\section{Mixed Automata}\label{s:automata}

Having introduced a notion of mixed strings, we now define a class of
deterministic automata that can accept mixed strings. Mixed
strings enforce a strict alternation between programs and tests, and
this alternation is reflected in our automata. The transitions of the 
automata are labelled with primitive programs and literals. Given a
mixed string, mixed automaton can process the tests in the string in 
many different orders; this reflects the fact that the tests that
appear in mixed strings are sets of literals.

A \emph{mixed automaton} over the set of primitive programs 
$\primprogs$ and set of primitive tests $\primtests$ is
a 3-tuple
$M=(\< S_A\>_{A\in\wp(\primtests)}, o, \<\delta_A\>_{A\in\wp(\primtests)})$,
consisting of 
a set $S_A$ of states for each possible base $A\ne\emptyset$ of a test
as well as a set $S_{\emptyset}$ of program states,
an output function $o : S_\emptyset \rightarrow \{ 0, 1 \}$, and
transition functions 
$\delta_\emptyset : S_\emptyset \times \primprogs \rightarrow S_\primtests$ 
and (for $A\neq\emptyset$) 
$\delta_A : S_A \times \lit(A) \rightarrow \bigcup_{A\in\wp(\primtests)}S_A$,
subject to the following two conditions:

\begin{enumerate}
\item[\condition{A1}.]
$\delta_A(s,l)\in S_{A\setminus\{\base(l)\}}$, and
\item[\condition{A2}.]
for every state $s$ in $S_A$, for every test
$t$ with base $A$, and for any two orderings $\<x_1,\ldots,x_m\>$,
$\<y_1,\ldots,y_m\>$ of the literals in $t$, if 
$s\trans{x_1}\ldots\trans{x_m}s_1$ and $s\trans{y_1}\ldots\trans{y_m}s_2$
then $s_1=s_2$. 

(For convenience, we write
$s\trans{l}s'$ if $\delta_A(s,l)=s'$ for $A$ the base of $s$.)
\end{enumerate}

We give an example of a mixed automaton in
Example~\ref{x:mixedautomaton}. 
Intuitively, a state in $S_A$ can process a mixed string of type
$A\rightarrow B$, for some $B$. Condition \condition{A1} enforces the
invariant that, as a string is being processed, the current state is
in $S_A$, for $A$ the base of the first element of the
string. Condition \condition{A2} is a form of ``path independence'':
regardless of the order in which we process the literals of a test, we
end up in the same program state. Condition \condition{A2}, and basing 
transitions on literals rather than tests, allow the manipulation of
mixed expressions at a finer level of granularity. This is related to 
a similar choice we made when allowing mixed strings to start with a
test that is not ``complete''. This flexibility will be useful when
we analyze mixed expressions in Section~\ref{s:expressions}.

The accepting states are defined via
the output function $o(s)$, viewed as a characteristic
function. Accepting states are in $S_\emptyset$. 

As in the coalgebraic treatment of automata \cite{Rutten98}, and
contrary to standard definitions, we allow both the state spaces $S_A$
and the set $\primprogs$ of primitive programs to be
infinite. We also do not force 
mixed automata to have initial states, for reasons that will become clear.
We now define the mixed language accepted by a state of a mixed
automaton. 
Call a sequence 
$\mu=e_1 \ldots e_m$
of primitive programs and literals a
\emph{linearization} of a mixed string
$\sigma = a_1 \ldots a_n$ if  $\mu$
can be obtained from $\sigma$ by 
replacing
each test $a_i$ in $\sigma$ with a sequence of
length $|a_i|$ containing exactly the literals in $a_i$.

\begin{example}
Let $\primprogs=\{p,q\}$, and $\primtests=\{b,c\}$. The mixed string 
 $\{ b \} p \{ \Neg{b}, c \} q \{ b, \Neg{c} \}$ 
 (of type $\{ b \} \rightarrow \emptyset$) has four linearizations:
 $b p \Neg{b} c q b \Neg{c}$, 
 $b p c \Neg{b} q b \Neg{c}$, 
 $b p \Neg{b} c q \Neg{c} b$, and 
 $b p c \Neg{b} q \Neg{c} b$. 
\end{example}

Intuitively, a mixed string $\sigma$ is accepted by an automaton
if a linearization of $\sigma$ is accepted by the automaton
according to the usual definition.
Formally, 
a mixed string $\sigma$ is \emph{accepted} by a state $s$
of an automaton $M$ 
if either 
\begin{enumerate}
\item $\sigma$ is $\epsilon$ and $s$ is a program
state with $o(s)=1$ (i.e., $s$ is an accepting program state), or
\item there exists a linearization $e_1 \ldots e_m$ of $\sigma$
such that $s \trans{e_1} \ldots \trans{e_m} s'$,
$s'$ is a program state,
and $o(s') = 1$.
\end{enumerate}
If $\sigma$ is accepted (by a state $s$) in virtue of satisfying the second
criterion, then every linearization is a witness to this fact---
in other words, the existential quantification in the second criterion
could be 
replaced with
a universal quantification (over all 
linearizations of $\sigma$) without any change in the actual
definition.  This is because of condition \condition{A2} in the
definition of a mixed automaton.
\begin{figure}[t]
\begin{center}
\includegraphics[width=4in]{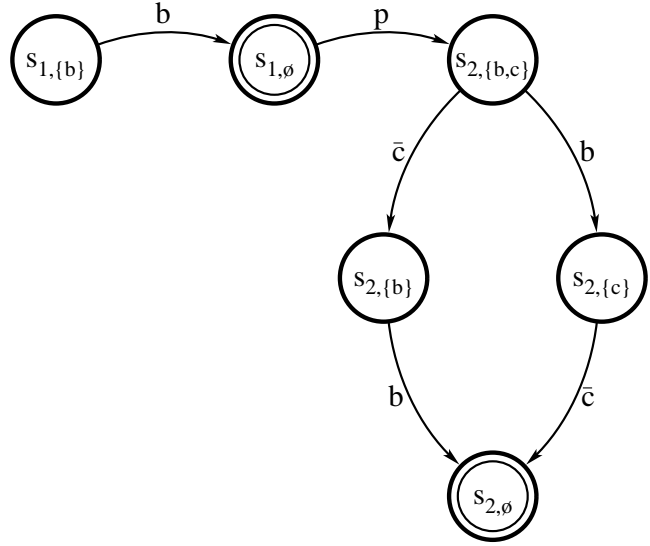}
\end{center}
\caption{A mixed automaton}
\label{f:aut}
\end{figure}

We define
the mixed language accepted by state $s$ of automaton $M$, written
$L_M(s)$, as the set of mixed strings accepted by state $s$ of $M$. It
is easy to verify that all the strings accepted by a state have the
same type, namely, if $s$ is in $S_A$, then every string in $L_M(s)$
has type $A\rightarrow\emptyset$, and hence $L_M(s)$ has type
$A\rightarrow\emptyset$. 

\begin{example}\label{x:mixedautomaton}
Let $\primprogs=\{p,q\}$, and $\primtests=\{b,c\}$. 
Consider the mixed automaton over $\primprogs$ and
$\primtests$ pictured in Figure~\ref{f:aut}, given by 
$M=(\<S_A\>_{A\in\wp(\primtests)},o,\<\delta_A\>_{A\in\wp(\primtests)})$, 
where: 
\begin{align*}
  & S_{\{b,c\}} = \{s_{2,\{b,c\}},s_{\text{sink},\{b,c\}}\}\\
  & S_{\{b\}} = \{s_{1,\{b\}},s_{2,\{b\}},s_{\text{sink},\{b\}}\}\\ 
  & S_{\{c\}} = \{s_{2,\{c\}},s_{\text{sink},\{c\}}\}\\ 
  & S_{\emptyset} = \{s_{1,\emptyset},s_{2,\emptyset},s_{\text{sink},\emptyset}\}\\
\intertext{and}
  & o(s_{1,\emptyset}) =  1\\ 
  & o(s_{2,\emptyset}) =  1\\ 
  & o(s_{\text{sink},\emptyset}) =  0.
\end{align*}
The transition function $\delta_{A}$ can be read off from
Figure~\ref{f:aut}; note that the sink states $s_{\text{sink},A}$ as
well as the transitions to the sink states are not
pictured. Intuitively, any transition not pictured in the automaton
can be understood as going to the appropriate sink state. For
instance, we have $\delta_{\{b,c\}}(s_{2,\{b,c\}},c) =
s_{\text{sink},\{b\}}$. 
We can check that the two conditions \condition{A1} and \condition{A2} 
hold in $M$. The language accepted by state $s_{1,\{b\}}$ is
$L_M(s_{1,\{b\}})=\{ \{b\},\{b\}p\{b,\Neg{c}\} \}$. The language
accepted by state $s_{1,\emptyset}$ is
$L_M(s_{1,\emptyset})=\{\epsilon,p\{b,\Neg{c}\}\}$. 
\end{example}

We define a \emph{homomorphism} between mixed automata $M$ and $M'$ to 
be a 
family 
$f=\<f_A\>_{A\in\wp(\primtests)}$ of functions
$f_A:S_A\rightarrow S'_A$ such that:
\begin{enumerate}
\item for all $s\in S_\emptyset$, $o(s)=o'(f_\emptyset(s))$, and
for all $p\in\primprogs$,
$f_\primtests(\delta_\emptyset(s,p))=\delta'_\emptyset(f_\emptyset(s),p)$, 
\item for all $s\in S_A$ (where $A\neq\emptyset$) and all $l\in\lit(A)$, 
$f_{A\setminus\{\base(l)\}}(\delta_A(s,l))=\delta'_A(f_A(s),l)$.

\end{enumerate}
A homomorphism preserves accepting states and transitions. 
We write $f:M\rightarrow M'$ when $f$ is a homomorphism between
automata $M$ and $M'$. For convenience, we often write $f(s)$ for
$f_A(s)$ when the type $A$ of $s$ is understood. 
It is straightforward to verify that mixed automata form a category
(denoted $\MA$), where the morphisms of the category are mixed
automata homomorphisms.

We are interested in identifying states that have the same behaviour,
that is, that accept the same mixed language. 
A \emph{bisimulation} between two mixed automata
$M=(\<S_A\>_{A\in\wp(\primtests)}, o, \<\delta_A\>_{A\in\wp(\primtests)})$
and
$M'=(\<S'_A\>_{A\in\wp(\primtests)}, o', \<\delta'_A\>_{A\in\wp(\primtests)})$
is a 
family
of relations $\<R_A\>_{A\in\wp(\primtests)}$ where
$R_A\subseteq S_A \times S'_A$ such that 
the following two conditions hold:
\begin{enumerate}

\item for all $s \in S_\emptyset$ and $s' \in S'_\emptyset$, 
if $s R_\emptyset s'$, then 
$o(s) = o'(s')$ and for all $p\in\primprogs$, 
$\delta_\emptyset(s,p) R_\primtests\delta'_\emptyset(s',p)$, and
\item for all $s \in S_A$ and $s' \in S'_A$ (where $A\neq\emptyset$), 
if $s R_A s'$, then 
for all $l\in\lit(A)$, 
$\delta_A(s,l) R_{A\setminus\{\base(l)\}} \delta'_A(s',l)$. 

\end{enumerate}

A bisimulation between $M$ and itself is called a bisimulation on
$M$. Two states $s$ and $s'$ of $M$ having the same type $B$ are said to be 
\emph{bisimilar}, denoted by
$s \sim_{M} s'$, if there
exists a bisimulation $\<R_A\>_{A\in\wp(\primtests)}$ such that $s R_B
s'$. (We simply write $s\sim s'$ when $M$ is clear from the context.)
For each $M$, the relation $\sim_M$ is the union of all bisimulations
on $M$, and in fact is the greatest bisimulation on $M$.

\begin{proposition}\label{p:bisim-states}
If $s$ is a state of $M$ and $s'$ is a state of $M'$ with $s\sim s'$,
then $L_M(s)=L_{M'}(s')$.
\end{proposition}
\begin{prf}
We show, by induction on the length of mixed strings that for all
mixed strings $\sigma$, and for all states $s,s'$ such that $s\sim
s'$, then $\sigma\in L_M(s)$ if and only if $\sigma\in
L_{M'}(s')$. For the empty mixed string $\epsilon$, we have
$\epsilon\in L_M(s)$ if and only if $o(s)=1$ if and only if $o'(s')=1$
(by definition of bisimilarity) if and only if $\epsilon\in
L_{M'}(s')$. Assume inductively that the results holds for mixed
strings of length $n$. Let $\sigma$ be a mixed string of length
$n+1$, of the form $a \sigma'$. Assume $\sigma\in L_M(s)$. By
definition, there is a linearization $e_1\ldots e_m$ of $a$ and a
state $s_1$ such that $s\trans{e_1}\ldots\trans{e_m}s_1$ and $\sigma'\in
L_M(s_1)$. By the definition of bisimilar states, we have
$s'\trans{e_1}\ldots\trans{e_m}s'_1$ and $s_1\sim s'_1$. By the
induction hypothesis, $\sigma'\in L_{M'}(s'_1)$. By the choice of
$s'_1$, we have that $\sigma\in L_{M'}(s')$, as desired. \qed
\end{prf}

Conditions (1) and (2) of the definition of a bisimulation are
analogous to the conditions in the definition of a
homomorphism. Indeed, a homomorphism can be viewed as a bisimulation.

\begin{proposition}\label{p:hom-bisim}
If $f:M\rightarrow M'$ is a mixed automataon homomorphism, then 
$\<R_A\>_{A\in\wp(\primtests)}$, defined by $R_A=\{(s,f_A(s))~:~s\in
S_A\}$ is a bisimulation.
\end{proposition}
\begin{prf}
First, for all $s\in S_\emptyset$, $s
R_\emptyset s'$ implies $s'=f_\emptyset(s)$, and
$o(s)=o'(f_\emptyset(s))=o'(s')$. Moreover, for all $p\in\primprogs$,
we have $\delta'_\emptyset(s',p) = \delta'_\emptyset(f_\emptyset(s),p)
= f_\primtests(\delta_\emptyset(s,l))$, so that $\delta_\emptyset(s,l)
R_\primtests \delta'_\emptyset(s',l)$, as required. Similarly, let
$s\in S_A$ 
(where $A\ne\emptyset$); 
$s R_A s'$ implies $s'=f_A(s)$, and thus
for all $l\in\lit(A)$,
$\delta'_A(s',l)=\delta'_A(f_A(s),l)=f_{A\setminus\{\base(l)\}}(\delta_A(s,l))$,
so that $\delta_A(s,l)R_{A\setminus\{\base(l)\}}\delta'_A(s',l)$, as
required, proving that $\<R_A\>_{A\in\wp(\primtests)}$ is a bisimulation. \qed
\end{prf}

An immediate consequence of this relationship is that homomorphisms
preserve accepted languages.

\begin{proposition}\label{p:hom-preserve-langs}
If $f:M\rightarrow M'$ is a mixed automaton homomorphism, then
$L_M(s)=L_{M'}(f(s))$ for all states $s$ of $M$. 
\end{proposition}
\begin{prf}
Immediate from Propositions~\ref{p:hom-bisim} and
\ref{p:bisim-states}. \qed
\end{prf}

It turns out that we can impose a mixed automaton structure on the set
of \emph{all} mixed languages with type $A\rightarrow\emptyset$. We take
as states mixed languages of type $A\rightarrow\emptyset$. A state is
accepting if the empty string $\epsilon$ is in the language. 
It remains to define the transitions between states;
we adapt the idea of Brzozowski derivatives \cite{Brzozowski64}.  Our
definition of derivative depends on whether we are taking the
derivative with respect to a program element or a literal.

If the mixed language $L$ has
type $\emptyset\rightarrow B$ and $p\in\primprogs$ is a primitive program,
define   
\[D_p(L)=\{\sigma ~:~ p\cdot\sigma\in L\}.\]

If the mixed language $L$ has 
type $A\rightarrow B$ (for $A\neq\emptyset$) and $l\in\lit(A)$ is a literal, then
\[D_l(L)=\{\sigma ~:~ \{l\}\cdot\sigma\in L\}.\]

Define $\langs_A$ to be the set of mixed languages of type
$A\rightarrow\emptyset$. Define $\langs$ to be
$(\<\langs_A\>_{A\in\wp(\primtests)}, o_\langs,
\<\delta_A\>_{A\in\wp(\primtests)})$, 
where $o_\langs(L)=1$ if $\epsilon\in L$, and $0$ otherwise; 
$\delta_\emptyset(L,p)=D_p(L)$;
and $\delta_A(L,l)=D_l(L)$, for $A\neq\emptyset$ and $l\in\lit(A)$. 
It is easy to verify that $\langs$ is indeed a mixed automaton. The
following properties of $\langs$ are significant.

\begin{proposition}\label{p:canonical-hom}
For a mixed automaton $M$ with states $\<S_A\>_{A\in\wp(\primtests)}$,
the maps $f_A:S_A\rightarrow\langs$ mapping a state $s$ in $S_A$ to
the language $L_M(s)$ form a mixed automaton homomorphism. 
\end{proposition}
\begin{prf}
We check the two conditions for the family
$\<f_A\>_{A\in\wp(\primtests)}$ to be a homomorphism. First, given
$s\in S_\emptyset$, $o(s)=1$ if and only if $\epsilon\in L_M(s)$,
which is equivalent to $o_\langs(f_\emptyset(s))=1$. Moreover, given
$p\in\primtests$, $f_{\primtests}(\delta_\emptyset(s,p)) =
L_M(\delta_\emptyset(s,p)) = \{ \sigma ~:~ p\cdot\sigma\in L_M(s)\} =
D_p(L_M(s)) = D_p(f_\emptyset(s))$, as required. Similarly, given $s\in
S_A$ (where $A\ne\emptyset$), and $l\in\lit(A)$,
$f_{A\setminus\{\base(l)\}}(\delta_A(s,l)) = L_M(\delta_A(s,l)) =
\{\sigma ~:~ \{l\}\cdot\sigma\in L_M(s)\} = D_l(L_M(s)) =
D_l(f_A(s))$, as required. \qed
\end{prf}

\begin{proposition}\label{p:lang-recognized-final}
For any mixed language $L$ in $\langs$, the mixed language accepted by
state $L$ in $\langs$ is $L$ itself, that is, $L_\langs(L)=L$. 
\end{proposition}
\begin{prf}
We prove by induction on the length of linearizations of $\sigma$ that
for all mixed strings $\sigma$, $\sigma\in L$ if and only if $\sigma
\in L_\langs(L)$. For the empty mixed string $\epsilon$, we have
$\epsilon\in L \Leftrightarrow o_\langs(L)=1 \Leftrightarrow
\epsilon\in L_\langs(L)$. For $\sigma$ of the form $p \sigma'$, we
have $\sigma=p\cdot \sigma'$, and thus
we have $p\cdot\sigma'\in L \Leftrightarrow \sigma'\in D_p(L)$, which
by 
the
induction hypothesis holds if and only if $\sigma'\in
L_\langs(D_p(L)) \Leftrightarrow \sigma'\in D_p(L_\langs(L))$ (because
$L_\langs$ is a mixed automaton homomorphism from $\langs$ to
$\langs$), which is just equivalent to $p\cdot\sigma'\in
L_\langs(L)$. 
For $\sigma$ with a linearization $l e_1\ldots e_m$,
letting $\sigma'$ denote a string with linearization 
$e_1 \ldots e_m$,
we have
$\sigma=\{l\}\cdot\sigma'$, and we can derive in an exactly similar
manner that $\{l\}\cdot\sigma'\in L \Leftrightarrow \sigma'\in D_l(L)
\Leftrightarrow \sigma'\in L_\langs(D_l(L)) \Leftrightarrow \sigma'\in
D_l(L_\langs(L)) \Leftrightarrow \{l\}\cdot\sigma'\in L_\langs(L)
\Leftrightarrow \sigma\in L_\langs(L)$. \qed
\end{prf}

These facts combine into the following fundamental property of
$\langs$, namely, that $\langs$ is a final automaton.
\begin{theorem}
$\langs$ is final in the category $\MA$, that is, for every mixed
automaton $M$, there is a unique homomorphism from $M$ to $\langs$.
\end{theorem}
\begin{prf}
Let $M$ be a mixed automaton. By Proposition~\ref{p:canonical-hom},
there exists a homomorphism $f$ from $M$ to the final automaton
$\langs$, mapping a state $s$ to the language $L_M(s)$ accepted by
that state. Let $f'$ be another homomorphism from $M$ to $\langs$. To
establish uniqueness, we need to show that for any state $s$ of
$M$, we have $f(s)=f'(s)$:
\begin{alignat*}{2}
f(s) & =  L_M(s) & & \mbox{(by definition of $f$)}\\
 & =  L_\langs(f'(s)) & & \mbox{(by  Proposition~\ref{p:hom-preserve-langs})}\\
 & =  f'(s) & & \mbox{(by Proposition~\ref{p:lang-recognized-final}).}
\end{alignat*}
Hence, $f$ is the required unique homomorphism. \qed
\end{prf}

The finality of $\langs$ gives rise to the following coinduction proof
principle for language equality, in a way which is by now standard
\cite{Rutten00}.  
\begin{corollary}\label{c:coinduction}
For two mixed languages $K$ and $L$ of type $A\rightarrow\emptyset$,
if $K\sim L$ then $K=L$.
\end{corollary}

In other words, to establish the equality of two mixed languages, it
is sufficient to exhibit a bisimulation between the two languages when
viewed as states of the final automaton $\langs$. In the following
sections, we will use this principle to analyze equality of languages
described by a typed form of $\KAT$ expressions.

\section{Pseudo-Bisimulations}\label{s:pseudo}

The ``path independence'' condition (\condition{A2}) in the definition
of a mixed automaton gives mixed automata a certain form of
redundancy. It turns out that due to this redundancy, we can define a
simpler notion than bisimulation that still lets us establish the
bisimilarity of states. 

A \emph{pseudo-bisimulation} (relative to the ordering
$b_1, \ldots, b_{|\primtests|}$ of the primitive tests in
$\primtests$) between two mixed automata
$M=(\<S_A\>_{A\in\wp(\primtests)}, o,
\<\delta_A\>_{A\in\wp(\primtests)})$ and
$M'=(\<S'_A\>_{A\in\wp(\primtests)}, o',
\<\delta'_A\>_{A\in\wp(\primtests)})$ is a 
family
of relations
$\<R_i\>_{i = 0, \ldots, |\primtests|}$ where $R_i\subseteq S_{A_i}
\times S'_{A_i}$ (with $A_i$ denoting $\{ b_j : j \leq i, j \in \{ 1,
\ldots, |\primtests| \} \}$) such that the following two conditions
hold:
\begin{enumerate}

\item for all $s \in S_\emptyset$ and $s' \in S'_\emptyset$, 
if $s R_0 s'$, then 
$o(s) = o'(s')$ and for all $p\in\primprogs$, 
$\delta_\emptyset(s,p) R_{|\primtests|} \delta'_\emptyset(s',p)$, and

\item for all $i = 1, \ldots, |\primtests|$, 
for all $s \in S_{A_i}$ and $s' \in S'_{A_i}$,
if $s R_i s'$, then 
for all $l\in\lit(b_i)$, 
$\delta_{A_i}(s,l) R_{i-1} \delta'_{A_i}(s',l)$. 

\end{enumerate}

The sense in which pseudo-bisimulation is weaker than a bisimulation
is that there need not be a relation for each element of
$\wp(\primtests)$. As the following theorem shows, however, we can
always complete a pseudo-bisimulation to a bisimulation. 
\begin{theorem}
\label{thm:pseudo-bisimulation}
\sloppy{If $\<R_i\>_{i = 0, \ldots, |\primtests|}$ is a pseudo-bisimulation
(relative to the ordering}
 $b_1, \ldots, b_{|\primtests|}$
 of the primitive tests in $\primtests$),
then there exists a bisimulation $\langle R'_A \rangle$
such that $R'_{A_i} = R_i$ for all $i = 0, \ldots, |\primtests|$
(with $A_i$ denoting $\{ b_j : j \leq i, j \in \{ 1, \ldots, |\primtests| \} \}$).
\end{theorem}
\begin{prf}
Let $\<R_i\>_{i=0,\ldots,|\primtests|}$ be a pseudo-bisimulation
(relative to the ordering on primitive tests $b_1,\ldots, b_{|\primtests|}$). We define a
family of relations $R'_A\subseteq S_A\times S'_A$ for each
$A\in\wp(\primtests)$, and show that it forms a bisimulation with the
required property. The proof relies on the path independence condition
\condition{A2} of mixed automata in a fundamental way. Given
$A\in\wp(\primtests)$, let $i(A)$ be the largest
$i\in\{1,\ldots,|\primtests|\}$ such that $\{b_1,\ldots,b_i\}\subseteq
A$, and let $c(A)$ be the relative complement of
$\{b_1,\ldots,b_{i(A)}\}$ defined by
$A\setminus\{b_1,\ldots,b_{i(A)}\}$. We say that a sequence of
literals $l_1,\ldots,l_k$ is \emph{exhaustive} over a set of bases $A$ 
if $A=\{\base(l_1),\ldots,\base(l_k)\}$ and $|A|=k$. 
Define $R'_A$ as follows: $s R'_A
s'$ holds if and only 
if for all literal sequences $l_1,\ldots,l_k$ exhaustive over $c(A)$, we
have  $s\trans{l_1}\ldots\trans{l_k} s_1$, 
$s'\trans{l_1}\ldots\trans{l_k} s'_1$, and $s_1 R_{i(A)}
s'_1$. Clearly, if $A=\{b_1,\ldots,b_{i(A)}\}$, then $R'_A=R_{i(A)}$,
as required. We now check that $\<R'_A\>_{A\in\wp(\primtests)}$ is a
bisimulation. Clearly, since $R'_\emptyset = R_0$, if $s R'_\emptyset
s'$, then $s R_0 s'$, and hence $o(s)=o'(s')$, and for all
$p\in\primprogs$, it holds that
$\delta_\emptyset(s,p)R_{|\primtests|}\delta_\emptyset(s',p)$, implying
$\delta_\emptyset(s,p)R'_\primtests\delta_\emptyset(s',p)$. Now, let
$A\neq\emptyset$, $s\in S_A$, $s'\in S'_A$, $l\in\lit(A)$, and assume
$s R'_A s'$. Consider the following cases:

Case $A=\{b_1,\ldots,b_{i(A)}\}, \base(l)=b_{i(A)}$: Since $s R'_A s'$, then
$s R_{i(A)} s'$, and by the properties of pseudo-bisimulations, we
have $\delta_A(s,l) R_{i(A)-1} \delta_A(s',l)$, which is exactly
$\delta_A(s,l) R'_{A\setminus\{\base(l)\}} \delta_A(s',l)$.

Case $A=\{b_1,\ldots,b_{i(A)}\}, \base(l)=b_j, j<i(A)$: Since $s R'_A
s'$, then $s R_{i(A)} s'$. let $l_1,\ldots,l_k$ be an arbitrary
exhaustive sequence of literals over
$\{b_{i(A)},\ldots,b_{j+1}\}$. Let $l'_{i(A)},\ldots,l'_{j+1}$ be the
arrangement of $l_1,\ldots,l_k$ such that $\base(l'_m)=b_m$.  Consider
the states $s_1,s_2,s'_1,s'_2$ such that
$s\trans{l'_{i(A)}}\ldots\trans{l'_{j+1}} s_1 \trans{l} s_2$, and
$s'\trans{l'_{i(A)}}\ldots\trans{l'_{j+1}} s'_1 \trans{l} s'_2$. By
the definition of pseudo-bisimulation, we have that $s_2 R_{j-1}
s'_2$. Now, by condition \condition{A2}, we have states $s_3,s'_3$
such that $s\trans{l}s_3\trans{l'_{i(A)}}\ldots\trans{l'_{j+1}}s_2$
and $s'\trans{l}s'_3\trans{l'_{i(A)}}\ldots\trans{l'_{j+1}}s'_2$. By
condition \condition{A2} again, we have that
$s_3\trans{l_1}\ldots\trans{l_k}s_2$ and
$s'_3\trans{l_1}\ldots\trans{l_k}s'_2$. Since $l_1,\ldots,l_k$ was
arbitrary, $s_2 R_{j-1} s'_2$ and $i(A\setminus\{\base(l)\})=j-1$, we
have $s_3 R'_{A\setminus\{\base(l)\}} s'_3$, that is, $\delta_A(s,l)
R'_{A\setminus\{\base(l)\}} \delta_A(s',l)$.

Case $A\supset\{b_1,\ldots,b_{i(A)}\}, \base(l)\in c(A))$: Pick an
arbitrary sequence $l_1,\ldots,l_k$ of literals that is exhaustive
over $c(A\setminus\{\base(l)\})$, and states $s_1,s_2,s'_1,s'_2$ such
that $s\trans{l}s_2\trans{l_1}\ldots\trans{l_k}s_1$, and
$s'\trans{l}s'_2\trans{l_1}\ldots\trans{l_k}s'_1$. By definition of
$R'_A$, we have $s_1 R_{i(A)} s'_1$. Since the sequence of literals
$l_1,\ldots,l_k$ was arbitrary, and since
$i(A)=i(A\setminus\{\base(l)\})$, we have that $s_2
R'_{A\setminus\{\base(l)\}} s'_2$, that is, $\delta_A(s,l)
R'_{A\setminus\{\base(l)\}} \delta_A(s',l)$.

Case $A\supset\{b_1,\ldots,b_{i(A)}\}, \base(l)=b_{i(A)}$: Pick an
arbitrary sequence $l_1,\ldots,l_k$ of literals that is exhaustive
over $c(A)$, and states $s_1,s'_1$ such that
$s\trans{l_1}\ldots\trans{l_k}s_1$ and
$s'\trans{l_1}\ldots\trans{l_k}s'_1$. By definition of $R'_A$, we have
$s_1 R_{i(A)} s'_1$. By definition of pseudo-bisimulation, if
$s_1\trans{l}s_2$ and $s'_1\trans{l}s'_2$, then we have $s_2
R_{i(A)-1} s'_2$. By condition
\condition{A2}, we have that for states $s_3,s'_3$,
$s\trans{l}s_3\trans{l_1}\ldots\trans{l_k}s_2$ and
$s'\trans{l}s'_3\trans{l_1}\ldots\trans{l_k} s'_2$. Thus, since
$l_1,\ldots,l_k$ was arbitrary, and 
$i(A\setminus\{\base(l)\})=i(A)-1$, we have $s_3
R'_{A\setminus\{\base(l)\}} s'_3$, that is, $\delta_A(s,l)
R'_{A\setminus\{\base(l)\}} \delta_A(s',l)$. 

Case $A\supset\{b_1,\ldots,b_{i(A)}\}, \base(l)=b_j, j<i(A)$: Pick an
arbitrary sequence $l_1,\ldots,l_k$ of literals that is exhaustive
over $c(A)\cup\{b_{i(A)},\ldots,b_{j+1}\}$. Let $l'_1,\ldots,l'_{k'}$
be the elements of $l_1,\ldots,l_k$ with bases in $c(A)$. Let
$l''_1,\ldots,l''_{k''}$ be the elements of $l_1,\ldots,l_k$ with
bases in $\{b_{i(A)},\ldots,b_{j+1}\}$. Let
$l'''_{i(A)},\ldots,l'''_{j+1}$ be the arrangement of
$l''_1,\ldots,l''_{k''}$ such that $\base(l'''_m)=b_m$. Consider
states $s_1,s'_1$ such that $s\trans{l'_1}\ldots\trans{l'_{k'}}s_1$
and $s'\trans{l'_1}\ldots\trans{l'_{k'}}s'_1$. By definition of
$R'_A$, we have $s_1 R_{i(A)} s'_1$. Now, consider states
$s_2,s_3,s'_2,s'_3$ such that
$s_1\trans{l'''_{i(A)}}\ldots\trans{l'''_{j+1}}s_2\trans{l}s_3$ and
$s'_1\trans{l'''_{i(A)}}\ldots\trans{l'''_{j+1}}s'_2\trans{l}s'_3$. By
the definition of pseudo-bisimulation, since $s_1 R_{i(A)} s'_1$, we
have that $s_3 R_{j-1} s'_3$. Now, by condition \condition{A2}, we
have states $s_4,s'_4$ such that
$s\trans{l}s_4\trans{l_1}\ldots\trans{l_k}s_3$ and
$s'\trans{l}s'_4\trans{l_1}\ldots\trans{l_k}s'_3$.  Since
$l_1,\ldots,l_k$ was arbitrary, and $i(A\setminus\{\base(l)\})=j-1$,
we have $s_4 R'_{A\setminus\{\base(l)\}} s'_4$, that is,
$\delta_A(s,l)R'_{A\setminus\{\base(l)\}}\delta_A(s',l)$. \qed

\end{prf}

Let us say that two states $s, s'$ are \emph{pseudo-bisimilar} if they
are related by some $R_i$ in a pseudo-bisimulation $\<R_i\>$; it
follows directly from Theorem~\ref{thm:pseudo-bisimulation} that
pseudo-bisimilar states are bisimilar.

\section{Mixed Expressions and Derivatives}\label{s:expressions}

A \emph{mixed expression} 
(over the set of primitive programs $\primprogs$ and
the set of primitive tests $\primtests$)
is any expression 
built via the following grammar:
\[
e ::= 0 ~~|~~ 1 ~~|~~ p ~~|~~ l ~~|~~ e_1 + e_2 ~~|~~ e_1 \cdot e_2 ~~|~~ e^{*}
\]
(with $p\in\primprogs$ and $l\in\lit(\primtests)$). For simplicity, we
often write $e_1 e_2$ for $e_1\cdot e_2$.  We also freely use
parentheses when appropriate.  Intuitively, the constants $0$ and $1$
stand for failure and success, respectively. The expression $p$
represents a primitive 
program,
while $l$ represents a primitive
test. The operation $+$ is used for choice, $\cdot$ for sequencing,
and $^*$ for iteration.  
These are a subclass of the KAT expressions
as defined by Kozen \citeyear{Kozen97}. (In addition to allowing
negated primitive tests, Kozen also allows negated tests.)  We call
them mixed expressions to emphasize the different interpretation we
have in mind.

In a way similar to regular expressions denoting regular languages, we
define a mapping $M$ from mixed expressions to mixed languages
inductively as follows: 
\begin{align*}
& M(0)  =   \emptyset\\
& M(1)  =   \{ \epsilon \}\\
& M(p)  =  \{ p \}\\
& M(l)  =  \{ \{l\} \}\\
&  M(e_1 + e_2)  =   M(e_1) \cup M(e_2)\\
&  M(e_1 \cdot e_2)  =   M(e_1) \cdot M(e_2)\\
&  M(e^*)  =   \bigcup_{n \geq 0} M(e)^n.
\end{align*}

The mapping $M$ is a rather canonical homomorphism from mixed
expressions to mixed languages.  (It is worth noting that we have not
defined any axioms for deriving the ``equivalence'' of mixed
expressions, and it is quite possible for distinct mixed expressions
to give rise to the same mixed language.)

Inspired by a type system devised by Kozen \citeyear{Kozen98,Kozen02}
for $\KA$ and $\KAT$ expressions, we impose a type system on mixed
expressions. The types have the form $A\rightarrow B$, where
$A,B\in\wp(\primtests)$, the same types we assigned to mixed
strings in Section~\ref{s:languages}. We shall soon see that this is
no accident. We assign a type to a mixed expression via a \emph{type
judgment} written $\vdash e : A\rightarrow B$. The following inference
rules are used to derive the type of a mixed expression:
\[ \vdash 0: A \rightarrow B \qquad
   \vdash 1: A \rightarrow A \qquad
   \vdash p: \emptyset \rightarrow B \]
\[
   \vdash l: A \cup \{ \base(l) \} \rightarrow A 
             \setminus \{ \base(l)\}\]

\[ \Rule{\vdash e_1: A \rightarrow B \quad \vdash e_2: A \rightarrow B}
       {\vdash e_1 + e_2: A \rightarrow B} \qquad
  \Rule{\vdash e_1: A \rightarrow B \quad \vdash e_2: B \rightarrow C}
       {\vdash e_1 \cdot e_2: A \rightarrow C}\]
\[
  \Rule{e: A \rightarrow A}{e^*: A \rightarrow A}.\]

It is clear from these rules that 
any subexpression of a mixed expression
having a type judgment also has a type judgment.

The typeable mixed expressions (which intuitively
are the ``well-formed'' expressions) induce typeable mixed languages
via the mapping $M$, as formalized by the following proposition.
\begin{proposition}\label{prop:types}
If $~\vdash e : A\rightarrow B$, then $M(e)$ is a mixed language of type
$A \rightarrow B$. 
\end{proposition}
\begin{prf}
A straightforward induction on the structure of mixed expressions. \qed
\end{prf}

Our goal is to manipulate mixed languages by manipulating the mixed
expressions that represent them via the mapping
$M$. (Of course, not every mixed language is in the image of $M$.) In
particular, we are interested in the operations $T(L)$ and
$\epsilon(L)$, as defined in Section~\ref{s:languages}, as well as the
language derivatives $D_p$ and $D_l$ introduced in the last section.

We now define operators on mixed expressions that capture those
operators on the languages denoted by those mixed expressions. We
define $\hT$ inductively on the structure of mixed expressions, as
follows: 
\begin{align*}
& \hT( 0 )  =   0\\
& \hT(1)  =   1\\
& \hT(p)  =   0\\
& \hT(l)  =   l\\
& \hT(e_1+e_2)  =  \hT(e_1)+\hT(e_2)\\
& \hT(e_1\cdot e_2)  =  \hT(e_1)\cdot\hT(e_2)\\
& \hT(e^*)  =  \hT(e)^*
\end{align*}
(where $p \in \primprogs$ and $l \in \lit(\primtests)$).  The operator
$\hT$ ``models'' the operator $T(L)$, as is made precise in the following
way.
\begin{proposition}
\label{prop:Tcommutes}
If $~\vdash e: A\rightarrow B$, 
then $\hT(e)$ is a typeable mixed expression such that
$T(M(e)) = M(\hT(e)).$
\end{proposition}
\begin{prf}
A straightforward induction on the structure of mixed expressions. \qed
\end{prf}

We define $\hepsilon$ inductively on the structure of mixed
expressions, as follows:
\begin{align*}
& \hepsilon( 0 ) =  0\\
& \hepsilon(1)  =  1\\
& \hepsilon(p)  =  0\\
& \hepsilon(l)  =  0\\
& \hepsilon(e_1+e_2)  =  
  \condarray{\hepsilon(e_1)=\hepsilon(e_2)=0}{0}{1}\\
& \hepsilon(e_1\cdot e_2)  =  
  \condarray{\hepsilon(e_1)=\hepsilon(e_2)=1}{1}{0}\\
& \hepsilon(e^*)  =  1
\end{align*}
(where $p \in \primprogs$ and $l \in \lit(\primtests)$). 
Note that $\hepsilon(e)$ is always the mixed expression $0$ or $1$. 
In analogy to Proposition \ref{prop:Tcommutes}, we have the following fact 
connecting the $\epsilon$ and $\hepsilon$ operators.
\begin{proposition}
\label{prop:Ecommutes}
If $~\vdash e : A\rightarrow B$,
then $\hepsilon(e)$ is a typeable mixed expression such that
$\epsilon(M(e)) = M(\hepsilon(e))$.
\end{proposition}
\begin{prf}
A straightforward induction on the structure of mixed expressions. \qed
\end{prf}

Finally, we define, by induction on the structure of mixed
expressions, the derivative operator $\hD$ for typeable mixed
expressions.  There are two forms of the derivative, 
corresponding
to the
two forms of derivative for mixed languages: the derivative $\hD_l$
with respect to a literal $l \in \lit(\primtests)$, and the derivative
$\hD_p$ with respect to a primitive program $p \in \primprogs$.  The
two forms of derivative are defined similarly, except on the product
of two expressions. (Strictly speaking, since the definition of the
derivative depends on the type of the expressions being
differentiated, $\hD$ should take type derivations as arguments rather
than simply expressions. To lighten the notation, we write $\hD$ as
though 
it took
mixed expressions as arguments, with the understanding
that the appropriate types are available.)

The derivative $\hD_p$ with respect to a primitive program
$p\in\primprogs$ is defined as follows:
\begin{align*}
& \hD_p(0)  =  0\\
& \hD_p(1)  =  0\\
& \hD_p(q)  =  \condarray{p=q}{1}{0}\\   %
& \hD_p(l)  =  0\\
& \hD_p(e_1 + e_2)  =  \hD_p(e_1) + \hD_p(e_2)\\
& \hD_p(e_1 \cdot e_2)  =  
    \condarray{B\neq\emptyset}{\hD_p(e_1) \cdot e_2}
        {\hD_p(e_1) \cdot e_2 + \hepsilon(e_1) \cdot \hD_p(e_2)}\\*
& \qquad \mbox{where $\vdash e_1:A\rightarrow B$ and $\vdash e_2:B\rightarrow C$}\\
& \hD_p(e^*)  =  \hD_p(e) \cdot e^*.
\end{align*}

The derivative $\hD_l$ with respect to a literal
$l\in\lit(\primtests)$ is defined as follows:
\begin{align*}
& \hD_l(0)  =  0\\
& \hD_l(1)  =  0\\
& \hD_l(p)  =  0\\
& \hD_l(l')  =  \condarray{l=l'}{1}{0}\\
& \hD_l(e_1 + e_2)  =  \hD_l(e_1) + \hD_l(e_2)\\
& \hD_l(e_1 \cdot e_2)  =  \condarray{\base(l)\notin B}{\hD_l(e_1)
\cdot e_2}{\hD_l(e_1) \cdot e_2 + \hT(e_1) \cdot \hD_l(e_2)}\\*
& \qquad \mbox{where $\vdash e_1:A\rightarrow B$ and $\vdash e_2:B\rightarrow C$}\\*
& \hD_l(e^*)  =  \hD_l(e) \cdot e^*.
\end{align*}

We have the following proposition, similar to the previous two, connecting
the derivative $\hD$ to the previously defined derivative $D$ on mixed languages.
\begin{proposition}
\label{prop:Dcommutes}
Suppose that $~\vdash e:A\rightarrow B$.

If $A=\emptyset$, then for all $p \in \primprogs$, $D_p(M(e)) = M(\hD_p(e)).$

If $A\neq\emptyset$, then for all $l\in\lit(A)$,
$D_l(M(e)) = M(\hD_l(e)).$
\end{proposition}
\begin{prf}
The proof is by induction on the structure of the mixed expression
$e$.  To illustrate the proof technique, we give one case of the
proof.

Suppose that $\vdash e_1: A \rightarrow B$ and $\vdash e_2: B
\rightarrow C$, and $e = e_1 \cdot e_2$.  Suppose further that $l \in
\lit(\primtests)$ is a literal such that $\base(l) \in A$ and
$\base(l) \in B$.  We will show that the proposition holds for the
expression $e$, assuming (by the induction hypothesis) that the
proposition holds for all subexpressions of $e$.

We first establish three claims that will be needed.

\noindent\textbf{Claim 1}: If $t$ is a test which (as a mixed string)
can be judged to have type $A \rightarrow B$,
then $\{ t \} \cdot \{ \sigma: \{ l \} \cdot \sigma \in M(e_2) \} =
\{ \sigma': \{ l \} \cdot \sigma' \in \{ t \} \cdot M(e_2) \}$.

First suppose that $\sigma$ is a mixed string 
such that $\{ l \} \cdot \sigma \in M(e_2)$.
Then $\sigma$
can be judged to have type 
$B \setminus \{ \base(l) \} \rightarrow C$, and so 
$\{ l \} \cdot \{ t \} \cdot \sigma =
 \{ t \} \cdot \{ l \} \cdot \sigma \in \{ t \} \cdot M(e_2)$.
It follows that 
$t \cdot \sigma \in \{ \sigma': \{ l \} \cdot \sigma' \in \{ t \} \cdot M(e_2) \}$.
For the other direction, suppose that
$\sigma'$ is a mixed string such that 
$\{ l \} \cdot \sigma' \in \{ t \} \cdot M(e_2)$.
Then there exists a mixed string $\tau \in M(e_2)$ such that
$\{ l \} \cdot \sigma' = \{ t \} \cdot \tau$.
Since $t$ can be judged to have type $A \rightarrow B$
and $\base(l) \in A \cap B$, $\base(l) \notin t$ and
there exists a mixed string $\sigma$ such that
$\{ l \} \cdot \sigma' = \{ t \} \cdot \tau = 
 \{ l \} \cdot \{ t \} \cdot \sigma$.
Thus $\sigma' = \{ t \} \cdot \sigma$ where 
$\{ l \} \cdot \sigma \in M(e_2)$.

\noindent\textbf{Claim 2}: If $\sigma$ is a mixed string such that
$l \cdot \sigma \in M(e_1)$, then
$l \cdot \sigma \in M(e_1) \setminus T(M(e_1))$.

This claim holds because $\{ l \} \cdot \sigma \in M(e_1)$
implies that $\sigma$ has type $A \setminus \{ \base(l) \} \rightarrow B$;
since $B \not\subseteq A'$, by the definition of the type
of a mixed string, $|\sigma| > 1$ and so 
$|\{ l \} \cdot \sigma| > 1$.

\noindent\textbf{Claim 3}: 
$\{ \sigma: \{ l \}  \cdot  \sigma \in M(e_1) \setminus T(M(e_1)) \} \cdot M(e_2) =  
\{ \sigma: \{ l \}  \cdot  \sigma \in (M(e_1) \setminus T(M(e_1))) \cdot M(e_2) \}$

The $\subseteq$ direction is straightforward.  For the $\supseteq$
direction, let $\sigma$ be a mixed string in the second set; then,
there exist strings $\tau_1 \in M(e_1) \setminus T(M(e_1))$ and
$\tau_2 \in M(e_2)$ such that
$\{ l \} \cdot \sigma = \tau_1 \cdot \tau_2$.
All strings in $M(e_1)$ have type $A \rightarrow B$; since
$\base(l) \in B$, there are no strings in $M(e_1)$ of length one
consisting of a primitive program, and so $|\tau_1| > 3$.
Hence $\sigma = \sigma' \cdot \tau_2$ for some mixed string $\sigma'$
such that $\{ l \} \cdot \sigma' \in M(e_1) \setminus T(M(e_1))$.

Using these three claims, we show that $D_l(M(e)) = M(\hD_l(e))$:
\begin{alignat*}{2}
 M&(\hD_l(e_1 \cdot e_2)) & \\
 & =  M(\hD_l(e_1) \cdot e_2 + \hT(e_1) \cdot \hD_l(e_2))  && \mbox{(by
 definition of $\hD_l$)}\\ 
 & =  M(\hD_l(e_1)) \cdot M(e_2) \cup M(\hT(e_1)) \cdot M(\hD_l(e_2))
  && \mbox{(by definition of $M$)}\\
 & =  D_l(M(e_1))   \cdot M(e_2) \cup M(\hT(e_1)) \cdot D(M(e_2))  && \mbox{(by induction hypothesis)}\\
 &  =  D_l(M(e_1))   \cdot M(e_2) \cup T(M(e_1))   \cdot D(M(e_2)) && \mbox{(by Proposition \ref{prop:Tcommutes})}\\
 & =  \{ \sigma: \{ l \}  \cdot  \sigma \in M(e_1) \} \cdot M(e_2)
 \cup \\ & \qquad   T(M(e_1))  \cdot  \{ \sigma : \{ l \} \cdot \sigma
\in M(e_2)) \}  && \mbox{(by definition of $D_l$)}\\
 & =  \{ \sigma: \{ l \}  \cdot  \sigma \in M(e_1) \} \cdot M(e_2)
 \cup \\ & \qquad \{ \sigma: \{ l \} \cdot \sigma \in T(M(e_1)) \cdot
 M(e_2)) \}  && \mbox{(by Claim 1)}\\ 
 & =  \{ \sigma: \{ l \}  \cdot  \sigma \in M(e_1) \setminus
 T(M(e_1)) \} \cdot M(e_2)   \cup \\ & \qquad  \{ \sigma: \{ l \} \cdot \sigma \in
 T(M(e_1)) \cdot M(e_2)) \} && \mbox{(by Claim 2)}\\
 & =  \{ \sigma: \{ l \}  \cdot  \sigma \in (M(e_1) \setminus
 T(M(e_1)))  \cdot M(e_2) \} \cup \\ & \qquad 
    \{ \sigma: \{ l \} \cdot \sigma \in T(M(e_1)) \cdot M(e_2)) \}  && 
\mbox{(by Claim 3)}\\
 & =  \{ \sigma: \{ l \}  \cdot  \sigma \in M(e_1) \cdot M(e_2) \}  \\
 & =  D_l(M(e_1) \cdot M(e_2))  && \mbox{(by definition of $D_l$)}\\
 & =  D_l(M(e_1  \cdot   e_2))  && \mbox{(by definition of $M$)}.
\end{alignat*}
The other cases are similar. \qed
\end{prf}

\section{Example}\label{s:example}

In this section, we use the notions of pseudo-bisimulation and the
coinduction proof principle (Corollary~\ref{c:coinduction}), along
with the derivative operator $\hD$, to prove the equivalence of two
mixed languages specified as mixed expressions.

Fix $\primprogs$ to be the set of primitive programs $\{ p, q \}$,
and $\primtests$ to be the set of primitive tests $\{ b, c \}$. Let
$[b]$ be a shorthand for $(b+\Neg{b})$.
Define $\alpha$ to be the mixed expression
\[(bp([b]cq)^* \Neg{c})^* \Neg{b}\]
and $\beta$ to be the mixed expression
\[bp([b]cq + b \Neg{c} p)^* \Neg{c} \Neg{b} + \Neg{b}.\]

Our goal is to prove that $\alpha$ and $\beta$ are equivalent, in the
sense that they induce the same language via the mapping $M$. In other
words, we want to establish that $M(\alpha)=M(\beta)$. 
This example demonstrates the equivalence of the 
program
\begin{verbatim}
  while b do {
    p;
    while c do q
  }
\end{verbatim}
and the program
\begin{verbatim}
  if b then {
    p;
    while b + c do
      if c then q else p
  }
\end{verbatim}
This equivalence is a component of the proof of the classical 
result that every \emph{while} program can be simulated
by a \emph{while} program with at most one while loop,
as presented by Kozen \citeyear{Kozen97}. 
We refer the reader there for more details.

There are a few ways to establish this equivalence. One is to rely on
a sound and complete axiomatization of the equational theory of
$\KAT$, and derive the equivalence of $\alpha$ and $\beta$
algebraically \cite{Kozen96}. Another approach is to first construct
for each expression an automaton that accepts the language it denotes,
and then minimize both automata \cite{r:kozen03}. Two expressions are
then equal if the two resulting automata are isomorphic.

In this paper, we describe a third approach, using the coinductive
proof principle for mixed languages embodied by
Corollary~\ref{c:coinduction}. Since the theory we
developed in Section~\ref{s:automata} applies only to mixed languages
of type $A\rightarrow\emptyset$, we verify that indeed we have $\vdash
\alpha:\{b\}\rightarrow\emptyset$ and $\vdash
\beta:\{b\}\rightarrow\emptyset$, so that, by
Proposition~\ref{prop:types}, $M(\alpha)$ and $M(\beta)$ are languages
of type $\{b\}\rightarrow\emptyset$.

We prove the equivalence of $\alpha$ and $\beta$ by showing that 
the mixed languages $M(\alpha)$ and $M(\beta)$
are pseudo-bisimilar, that is, they are related by some
pseudo-bisimulation. More specifically, we exhibit a
pseudo-bisimulation, relative to the ordering $b_1 = b$, $b_2 = c$, on
the final automaton $\langs$, such that $M(\alpha)$ and $M(\beta)$ are
pseudo-bisimilar.  This is sufficient for proving equivalence, since
by Theorem~\ref{thm:pseudo-bisimulation}, the languages $M(\alpha)$
and $M(\beta)$ are then bisimilar, and by
Corollary~\ref{c:coinduction}, $M(\alpha)=M(\beta)$.

Define $\alpha'$ to be the mixed expression
\[([b]cq)^* \Neg{c} \alpha\]
and define $\beta'$ to be the mixed expression
\[([b]cq + b \Neg{c} p)^* \Neg{c} \Neg{b}.\] Notice that $\beta = bp
\beta' + \Neg{b}$.

We note that (using the notation of the definition
of pseudo-bisimulation), $A_0 = \emptyset$, $A_1 = \{ b \}$, and $A_2
= \{ b, c \}$.  We claim that the following three relations form a
pseudo-bisimulation:
\[ R_2 = \{ \begin{prog}
             (M(\alpha'), M(\beta')),\\
             (M(0),M(0)) \}\end{prog} \qquad
R_1 = \{ \begin{prog}
             (M([b]q \alpha'), M([b]q \beta')), \\
             (M(\alpha), M(\beta)) \}\end{prog}
\]
\[ R_0 = \{ \begin{prog}
             (M(p \alpha'), M(p \beta')),\\
             (M(q \alpha'), M(q \beta')),\\
             (M(1),M(1)),\\
             (M(0),M(0)) \}.\end{prog}\]

It is straightforward to verify that $\< R_0, R_1, R_2 \>$ is a 
pseudo-bisimulation on $\langs$,
using the operators defined in the previous section. For instance,
consider $D_b(M(\alpha))$, which is equal to $M(\hD_b(\alpha))$ by
Proposition~\ref{prop:Dcommutes}. 
We compute 
$\hD_{b}(\alpha)$ here:
\begin{align*}
\hD_{b}(\alpha) & =  
  \hD_b( (bp([b]cq)^* \Neg{c})^* ) \Neg{b} + \hT( (bp([b]cq)^*
  \Neg{c})^* ) \hD_b(\Neg{b}) \\
& =  \hD_b(  bp([b]cq)^* \Neg{c}    )(bp([b]cq)^* \Neg{c})^* \Neg{b}
  + \hT( (bp([b]cq)^* \Neg{c})^* ) 0 \\
& =       p([b]cq)^* \Neg{c}     (bp([b]cq)^* \Neg{c})^* \Neg{b}\\
& =       p \alpha'.
\end{align*}
Hence, $D_b(M(\alpha)) = M(\hD_b(\alpha)) = M(p\alpha')$. The other
cases are similar. 

As we shall see shortly, there is a way to mechanically construct such
a bisimulation to establish the equivalence of two mixed expressions.

We remark that an alternative approach to establish equivalence of
while programs based on coalgebras is described by Rutten
\citeyear{Rutten99}. This approach uses the operational semantics of the
programs instead of an algebraic framework.

\section{Completeness}\label{s:completeness}

Thus far, we have established a coinductive proof technique for
establishing the equality of mixed languages
(Section~\ref{s:automata}), and illustrated its use by showing the
equality of two particular mixed languages specified by mixed
expressions (Section~\ref{s:example}), making use of the derivative
calculus developed in Section~\ref{s:expressions}.  A natural
question about this proof technique is whether or not it can establish
the equivalence of
\emph{any} two mixed expressions that are equivalent
(in that they specify the same mixed language).
In this section, we answer this question in the affirmative
by formalizing and proving a completeness theorem for our proof
technique.
In particular, we show that 
given two equivalent mixed expressions,
a finite bisimulation relating them can be
effectively constructed, 
by performing only simple syntactic manipulations. 
In fact, we exhibit a deterministic procedure
for deciding whether or not two mixed expressions are equivalent.

In order to state our completeness theorem, we need a few definitions. We say
that
two mixed expressions $e_1$ and $e_2$ are \emph{equal up to ACI
properties}, written $e_1\eqACI e_2$, if  $e_1$ and $e_2$ 
are syntactically equal, up to the associativity, commutativity, and
idempotence of $+$. 
That is, $e_1$ and $e_2$ are equal up to ACI properties if the following
three rewriting rules can be applied to subexpressions of $e_1$ 
to obtain $e_2$:
\begin{gather*}
e + (f + g) = (e + f) + g\\
e + f = f + e\\
e + e = e.
\end{gather*}

Given a relation $\hR$
between mixed expressions, we define an induced relation 
$\hRACI$ as follows:
$e_1 \hRACI e_2$
 if and only if there exists $e_1',e_2'$ such that $e_1\eqACI e_1'$,
$e_2\eqACI e_2'$, and $e_1'\hR e_2'$. 

We define a \emph{syntactic bisimulation} between two mixed
expressions $e_1$ and $e_2$ having the same type $B\rightarrow\emptyset$ 
(for some $B \subseteq \primtests$)
to be a
family
$\hR=\<\hR_A\>_{A\in\wp(\primtests)}$ of relations such
that 
\begin{enumerate}
\item for all mixed expressions $e,e'$, if $e \hR_A e'$, then
$\vdash e:A\rightarrow\emptyset$ and $\vdash
e':A\rightarrow\emptyset$,
\item $e \hR_B e'$,
\item for all mixed expressions $e,e'$,
if $e \hR_\emptyset e'$,
then $\hepsilon(e) = \hepsilon(e')$, and for all $p\in\primprogs$,
$\hD_p(e)\hRACI_\primtests \hD_p(e')$, and
\item for all mixed expressions $e,e'$, if $e \hR_A e'$
(for $A\not=\emptyset$), then for all $l\in\lit(A)$, $\hD_l(e)
\hRACI_{A\backslash\{\base(l)\}} \hD_l(e')$. 
\end{enumerate}

A syntactic bisimulation resembles a bisimulation, but is
defined over mixed expressions, rather than over mixed languages. The
next theorem shows that any two equivalent mixed expressions are
related by a \emph{finite} syntactic bisimulation, that is, a
syntactic bisimulation $\hR$ where the number of pairs in each
relation $\hR_A$ is finite.

\begin{theorem}
\label{thm:syntactic-bisimulation}
For all mixed expressions $e_1,e_2$, of type $A\rightarrow\emptyset$,
$M(e_1)=M(e_2)$ if and only if there exists a finite syntactic
bisimulation between $e_1$ and $e_2$. 
\end{theorem}
\begin{prf}
$(\Leftarrow)$ It is easy to check that a syntactic bisimulation $\hR$
induces a bisimulation $R$ such that $e_1 \hR_A e_2$ if and only if
$M(e_1) R_A M(e_2)$. The result then follows by
Corollary~\ref{c:coinduction}. 

$(\Rightarrow)$ We first show how to construct, for every mixed
expression $e$ with $\vdash e:A_e\rightarrow B_e$, a finite-state
automaton
$M=(\<S_A\>_{A\in\wp(\primtests)},\<\delta_A\>_{A\in\wp(\primtests)})$ 
with transition functions
$\delta_\emptyset : S_\emptyset \times \primprogs \rightarrow S_\primtests$ 
and (for $A\neq\emptyset$) 
$\delta_A : S_A \times \lit(A) \rightarrow
\bigcup_{A\in\wp(\primtests)}S_A$, satisfying the conditions 
\begin{enumerate}
\item $\delta_A(s,l)\in S_{A\setminus\{\base(l)\}}$,
\item the states of $S_A$ are mixed expressions having type $A\rightarrow B_e$, 
\item $e$ is a state of $S_{A_e}$, 
\item if $\delta_{\emptyset}(s_1,p)=s_2$, then $\hD_p(s_1)\eqACI s_2$, and
\item if $\delta_{A}(s_1,l)=s_2$, then $\hD_l(s_1)\eqACI s_2$. 
\end{enumerate}
We define the automaton by induction on the structure of
$e$. 
The cases for $0,1,p,l$ 
are straightforward.
We focus on the remaining cases:  

Case $e=e_1+e_2$: Assume by induction that we have automata
$M_1$, $M_2$ for $e_1$ and $e_2$. Define:
\begin{align*}
& S_A=\{f_1+f_2~:~f_1\in S_{1,A},f_2\in S_{2,A}\}\\
& \delta_\emptyset(f_1+f_2,p)=\delta_\emptyset(f_1,p)+\delta_\emptyset(f_2,p)\\
& \delta_A(f_1+f_2,l)=\delta_{1,A}(f_1,l)+\delta_{2,A}(f_2,l), \mbox{for
$A\not=\emptyset,l\in\lit(A)$.}
\end{align*}

Case $e=e_1\cdot e_2$: Let $\vdash e_1:A_1\rightarrow B_1$. Assume
by induction that we have  
automata $M_1$,$M_2$ for $e_1$ and $e_2$. Define:
\begin{align*}
& S_A=\{\begin{prog}
      f\cdot e_2+\sum\limits_{(t,g)\in E}t\cdot g+\sum\limits_{g\in G}g~:~ \\
      \quad f\in S_{1,A},E\subseteq\mathit{Tests}(A \rightarrow
  B_1)\times S_{2,B_1},G\subseteq S_{2,A}\}
      \end{prog}\\
& \quad \begin{prog}
  \delta_\emptyset(f\cdot e_2+\sum\limits_{g\in G}g,p)=\\
  \quad\condarray{B=\emptyset,
  \hepsilon(f)=1}{\delta_{1,\emptyset}(f,p)\cdot e_2 +
  \delta_{2,\emptyset}(e_2,p)+\sum\limits_{g\in
  G}\delta_{2,\emptyset}(g,p)}{\delta_{1,\emptyset}(f,p)\cdot e_2 +
  \sum\limits_{g\in G}\delta_{2,\emptyset}(g,p)}
\end{prog}\\
& \quad \begin{prog}
 \delta_A(f\cdot e_2+\sum\limits_{(t,g)\in E}t\cdot g+\sum\limits_{g\in
  G}g,l) = \\
  \quad \left\{\begin{array}{ll}
    \delta_{1,A}(f,l) \cdot e_2 + \sum\limits_{(t,g)\in E}D_l(t)\cdot
g + \sum\limits_{g\in G}\delta_{2,A}(g,l) & \mbox{if $\base(l)\in
A\setminus B_1$}\\
    \delta_{1,A}(f,l) \cdot e_2 + \sum\limits_{(t,g)\in E}t\cdot
       \delta_{2,B_1}(g,l) + \sum\limits_{g\in G}\delta_{2,A}(g,l) &
       \mbox{if $\base(l)\not\in A\cup B_1$}\\
   \begin{prog}
     \delta_{1,A}(f,l)\cdot e_2 + \hT(f)\cdot\delta_{2,B_1}(e_2,l) + \\
     \quad  \sum\limits_{(t,g)\in E}t\cdot \delta_{2,B_1}(g,l) +
       \sum\limits_{g\in G}\delta_{2,A}(g,l)
   \end{prog} & 
       \mbox{if $\base(l)\in B_1$}\end{array}\right.\\
  \quad\mbox{for $A\not=\emptyset,l\in\lit(A)$.}
\end{prog}
\end{align*}

Case $e=e_1^*$: Let $\vdash e_1:A_1\rightarrow A_1$. Assume by
induction that we have an automaton  $M_1$ for $e_1$. Define:
\begin{align*}
& S_A=  \condarray{A= A_1}{\{\gamma\cdot e_1^*+ \sum\limits_{f\in F}f\cdot e_1^*~:~
   \gamma\in\{0,1\},F\subseteq S_{1,A_1}\}}{\{\sum\limits_{f\in
   F}f\cdot e_1^*~:~ F\subseteq S_{1,A}\}}\\
& \begin{prog}
 \delta_\emptyset(\gamma\cdot e_1^*+\sum\limits_{f\in F}f\cdot e_1^*,p) 
   =\\
  \quad 
      \gamma\cdot\delta_{1,\emptyset}(e_1,p)\cdot e_1^* +
       \sum\limits_{f\in F}\delta_{1,\emptyset}(f,p)\cdot e_1^* +
       \quad \sum\limits_{f\in
        F}\hepsilon(f)\cdot\delta_{1,\emptyset}(e,p)\cdot e_1^*,\\ 
   \qquad \mbox{for $A=A_1$}
     \end{prog}\\
& \delta_\emptyset(\sum\limits_{f\in F}f\cdot e_1^*,p)=
  \sum\limits_{f\in F}\delta_{1,\emptyset}(f,p)\cdot e_1^*, \mbox{for
  $A\not= A_1$},\\
& \begin{prog}
 \delta_A(\gamma\cdot e_1^*+\sum\limits_{f\in F}f\cdot e_1^*,l) =\\
 \quad\gamma\cdot\delta_{1,A}(e_1,l)\cdot e_1^* + \sum\limits_{f\in
  F}\delta_{1,A}(f,l)\cdot e_1^* + \sum\limits_{f\in F}\hepsilon(f)\cdot 
  \delta_{1,A}(e,l)\cdot e_1^*,\\
  \qquad\mbox{for $A\not=\emptyset,A=A_1,l\in\lit(A)$}\end{prog}\\
& \delta_A(\sum\limits_{f\in F}f\cdot e_1^*,l) = \sum\limits_{f\in
  F}\delta_{1,A}(f,l)\cdot e_1^*, \mbox{for $A\not=\emptyset,A\not=
A_1,l\in\lit(A)$.}
\end{align*}

It is straightforward (if tedious) to verify that the resulting
automaton satisfies
properties (1)-(5) given above.

This completes the construction of the finite state mixed
automaton corresponding to $e$.  

Given equivalent mixed expressions $e_1$ and $e_2$ of type
$A\rightarrow\emptyset$, a finite syntactic bisimulation $\hR$ can be
constructed as follows. First, construct the automata $M_1$ and $M_2$
corresponding to $e_1$ and $e_2$. Then, initialize $\hR$ to contain
the pair $(e_1,e_2)$, and iterate the following process: for every
$(e,e')$ in $\hR$, add the pairs
$(\delta_{1,B}(e,x),\delta_{2,B}(e',x))$ (where $e,e'$ have type 
$B\rightarrow\emptyset$), for all $x$.  Perform this iteration until
no new pairs are added to $\hR$. This must terminate, because there
are finitely many pairs of states $(e,e')$ with $e$ in $M_1$
and $e'$ in $M_2$. It is straightforward to check that $\hR$ is a
syntactic bisimulation, under the assumption that $M(e_1)=M(e_2)$. \qed 
\end{prf}

The procedure described in the proof of
Theorem~\ref{thm:syntactic-bisimulation}  
can in fact
be easily turned 
into a procedure for deciding if two mixed expressions are
equivalent. To perform this decision, construct $\hR$, and verify that 
at all pairs of states $(e,e')$ in $\hR$,
$\hepsilon(e)=\hepsilon(e')$. If this verification fails, then the two 
mixed expressions are not equivalent; otherwise, they are equivalent. 

The bisimulation in Section~\ref{s:example} is indeed a bisimulation
induced by a syntactic bisimulation on the mixed expressions
$\alpha$ and $\beta$.

\section{Conclusions and Future Work}\label{s:conclusions}

We believe that proofs of equivalence between mixed expressions such
as $\alpha$ and $\beta$ via bisimulation are in general more easily
derived than ones obtained through a sound and complete axiomatization
of $\KAT$. Given two equivalent mixed expressions, we can exhibit a
bisimulation using the purely mechanical procedure underlying
Theorem~\ref{thm:syntactic-bisimulation}: use the derivative operators
to construct a finite bisimulation in which the two expressions
are paired. In contrast, equational reasoning typically requires
creativity.

The ``path independence'' of a mixed automaton (condition
\condition{A2}) gives any mixed automaton a certain form of
redundancy.  This redundancy persists in the definition of
bisimulation, and is the reason why a pseudo-bisimulation, a seemingly
weaker notion of bisimulation, gives rise to a bisimulation.  An open
question is to cleanly eliminate this redundancy; a particular
motivation for doing this would be to make proofs of expression
equivalence as simple as possible.  Along these lines, it would be of
interest to develop other weaker notions of bisimulation that give
rise to bisimulations; pseudo-bisimulations require a sort of ``fixed
variable ordering'' that does not seem absolutely necessary.

Another issue for future work would be to give a class of expressions
wider than our mixed expressions for which there are readily
understandable and applicable rules for computing derivatives.  In
particular, a methodology for computing derivatives of the $\KAT$
expressions defined by Kozen \citeyear{Kozen97} would be nice to see.
Intuitively, there seems to be a tradeoff between the expressiveness
of the regular expression language and the simplicity of computing
derivatives (in the context of $\KAT$).  Formal 
work towards
understanding this tradeoff could potentially be quite useful.

\subsection*{Acknowledgments}

The authors wish to thank Dexter Kozen for helpful comments on a draft
of this paper,
as well as the anonymous referees who helped improve the presentation
of the results.
The second author was supported by NSF under grant
CTC-0208535, by ONR under grants N00014-00-1-03-41 and
N00014-01-10-511, and by the DoD Multidisciplinary University Research
Initiative (MURI) program administered by the ONR under grant
N00014-01-1-0795.

\bibliographystyle{chicagor}
\bibliography{riccardo,riccardo2}

\end{document}